\documentclass{emulateapj}

\newcommand\curf{{\cal F}}

\shorttitle{Accretion variability of Herbig Ae/Be stars observed by X-Shooter}
\shortauthors{Mendigut\'\i{}a et al.}

\begin{document}


\title{Accretion variability of Herbig Ae/Be stars observed by X-Shooter\\
    HD 31648 and HD 163296}


\author{I. Mendigut\'\i{}a}
\affil{Department of Physics and Astronomy, Clemson University, Clemson, SC 29634-0978, USA}
\email{imendig@clemson.edu}

\author{S.D. Brittain}
\affil{Department of Physics and Astronomy, Clemson University, Clemson, SC 29634-0978, USA}

\author{C. Eiroa}
\affil{Departamento de F\'{\i}sica Te\'{o}rica, M\'{o}dulo 15, Facultad de Ciencias, Universidad Aut\'{o}noma de Madrid, PO Box 28049, Cantoblanco, Madrid, Spain.}

\author{G. Meeus}
\affil{Departamento de F\'{\i}sica Te\'{o}rica, M\'{o}dulo 15, Facultad de Ciencias, Universidad Aut\'{o}noma de Madrid, PO Box 28049, Cantoblanco, Madrid, Spain.}

\author{B. Montesinos}
\affil{Centro de Astrobiolog\'{\i}a, Departamento de Astrof\'{\i}sica (CSIC-INTA), ESAC Campus, P.O. Box 78, 28691 Villanueva de la Ca\~nada, Madrid, Spain.}

\author{A. Mora}
\affil{GAIA Science Operations Centre, ESA, European Space Astronomy Centre, PO Box 78, 28691, Villanueva de la Ca\~nada, Madrid, Spain.}

\author{J. Muzerolle}
\affil{Space Telescope Science Institute, 3700 San Martin Dr., Baltimore, MD, 21218, USA}

\author{R.D. Oudmaijer}
\affil{School of Physics \& Astronomy, University of Leeds, Woodhouse Lane, Leeds LS2 9JT, UK}

\author{E. Rigliaco}
\affil{Department of Planetary Science, Lunar and Planetary Lab, University of Arizona, 1629, E. University Blvd, 85719, Tucson, AZ, USA}





\begin{abstract}
This work presents X-Shooter/VLT spectra of the prototypical, isolated Herbig Ae stars HD 31648 (MWC 480) and HD 163296 over five epochs separated by timescales ranging from days to months. Each spectrum spans over a wide wavelength range covering from 310 to 2475 nm. We have monitored the continuum excess in the Balmer region of the spectra and the luminosity of twelve ultraviolet, optical and near infrared spectral lines that are commonly used as accretion tracers for T Tauri stars. The observed strengths of the Balmer excesses have been reproduced from a magnetospheric accretion shock model, providing a mean mass accretion rate of 1.11 $\times$ 10$^{-7}$ and 4.50 $\times$ 10$^{-7}$ M$_{\odot}$ yr$^{-1}$ for HD 31648 and HD 163296, respectively. Accretion rate variations are observed, being more pronounced for HD 31648 (up to 0.5 dex). However, from the comparison with previous results it is found that the accretion rate of HD 163296 has increased by more than 1 dex, on a timescale of $\sim$ 15 years. Averaged accretion luminosities derived from the Balmer excess are consistent with the ones inferred from the empirical calibrations with the emission line luminosities, indicating that those can be extrapolated to HAe stars. In spite of that, the accretion rate variations do not generally coincide with those estimated from the line luminosities, suggesting that the empirical calibrations are not useful to accurately quantify accretion rate variability. 
\end{abstract}




\vspace*{5mm}

\section{Introduction}
\label{Sect:intro}
The star formation process can be studied from the evolution of the accretion rate \citep[see e.g.][]{Hartmann98,Siciliaguilar04,Fedele10}. At the initial stages, protostars show both envelope-to-disk and disk-to-star accretion, which can show variations of several orders of magnitude on relatively short timescales \citep[FUor and EXor outbursts, e.g.][]{HartmannKen96,Herbig08,Liskowsky10}. During the pre-main sequence (PMS) phase, once the envelope dissipates and the central object becomes optically visible, the mass accretion rate ($\dot{M}_{\rm acc}$) decreases to typical values of $\sim$ 10$^{-8}$ and 10$^{-7}$ M$_{\odot}$ yr$^{-1}$ for T Tauri (TT) and Herbig Ae/Be (HAeBe) stars, respectively \citep{Hartmann_98,Mendi11b}. The accretion rate variability in the PMS phase is also expected to be lower than in previous stages. However, the precise strength of those variations is not well known, which can be partly attributed to discrepancies measuring accretion.\\

Primary signatures of accretion are the excess of UV/optical continuum emission and the veiling in the spectroscopic absorption lines that are observed in accreting stars. Both can be modeled from hot accretion shocks whose emission superimposes to that of the stellar atmosphere \citep{CalvetGullbrig98}. The way that material from the inner disk reaches the stellar surface is understood from the magnetospheric accretion (MA) scenario, according to which gas does not fall directly from the disk to the star but follows the magnetic field lines that connect them \citep{Uchida85,Konigl91,Shu94}. This is the accepted view for TT stars and seems to be valid at least for several HAeBes \citep[see e.g.][]{Vink02,Muzerolle04,Mottram07,DonBrit11,Mendi13}. Accretion rates obtained from MA shock modeling correlate with the luminosity of several emission lines that span from the ultraviolet (UV) to the infrared (IR) \citep[see e.g. the compilation in][]{Rigliaco11}. These types of empirical calibrations have been extended to late-type HAeBes for the [\ion{O}{1}]6300, H$\alpha$ \citep{Mendi11b} and Br$\gamma$ \citep{DonBrit11,Mendi11b} lines. The origin of the correlations is not fully understood, but these spectral lines can be considered as secondary signatures of accretion, being useful to easily derive accretion rates for wide samples of stars.\\

Although some works are based on observed variations of primary accretion signatures in TT stars \citep[e.g.][]{Herbst94,FerEiroa96,Batalha01}, studies of accretion rate variability are mainly based on secondary ones \citep[e.g.][]{Nguyen09,Eisner10,Costigan12,Chou13}. These analyses provide different results depending on the spectroscopic line used as accretion tracer. As an example, \citet{Nguyen09} studied a sample of 40 TT stars, reporting typical accretion rate variations of 0.35 dex from the CaII8862 line flux and 0.65 dex from the H$\alpha$ 10$\%$. Similar results were more recently obtained by \citet{Costigan12}. In fact, differences in the variability shown by several lines used as accretion tracers are reported for both TTs and HAeBes \citep{JohnsBasri95,LagoGameiro98,Mendi11a}. However, accretion variability from a primary accretion signature was found to be typically lower than 0.5 dex for the HAeBes, on timescales from days to months \citep{Mendi11b}, a result confirmed by \citet{Pogodin12}. Accretion variability of specific TT stars, determined from line veiling, is also expected to be low \citep{Alencar02}. However, the simultaneous analysis of primary and secondary accretion signatures is necessary to test the reliability of the empirical calibrations between the accretion rate and different spectroscopic lines \citep{Rigliaco12}, and to study the validity of MA in early-type HAeBe stars \citep{DonBrit11,Mendi11b}. These types of studies have been boosted by the advent of the {\it X-Shooter} spectrograph at the {\it Very Large Telescope} \citep[VLT;][]{Vernet11}, which allows simultaneous coverage of the accretion signatures from the UV to the near-IR. Simultaneity is crucial when dealing with PMS stars, given their variable nature. Several published and ongoing works face those type of questions, by ''X-Shooting'' wide samples of PMS stars \citep{Oudmaijer11,Rigliaco12,Manara13}.\\              

This work aims to contribute to the understanding of the spectroscopic accretion tracers in HAeBe stars. Instead of relating primary and secondary accretion signatures from single-epoch spectra of large samples, we will analyze multi-epoch spectra of two relevant objects. This approach brings the additional possibility of analyzing the variability of the accretion rate. This paper focuses on the variability behavior of two HAe stars: \object{HD 31648} (MWC 480) and \object{HD 163296}. The specific objective is twofold. First, to measure the accretion rate and its variability from the Balmer excess \citep[the primary accretion signature for HAeBe stars, see][]{Garrison78,Muzerolle04}. Second, to compare the previous results with several UV, optical and NIR line luminosities commonly used as secondary accretion tracers in low-mass stars. This comparison will provide relevant information about the applicability of the spectroscopic accretion tracers in late-type HAeBe stars.\\

The paper is organized as follows: Sect. \ref{Sect:observs} describes the observations and data reduction, Sect. \ref{Section:results} presents our observational results on the Balmer excess (Sect. \ref{Subsection: varbalmer}) and the spectroscopic lines (Sect. \ref{Subsection: varlines}), Sect. \ref{Sect:analysis} includes the analysis of the Balmer excesses in terms of a MA model (Sect. \ref{Subsection: accretion_from_Balmer}), and discusses possible relations between the accretion luminosity derived from the Balmer excess and from the emission line luminosities (Sects. \ref{Subsection: accretion_lines} and \ref{Subsection: var_lont}). Finally, Sect. \ref{Sect:conclusions} summarizes our main conclusions.

\section{Stars, observations and data reduction}
\label{Sect:observs}

\begin{figure*}
\centering
 \includegraphics[width=175mm,clip=true]{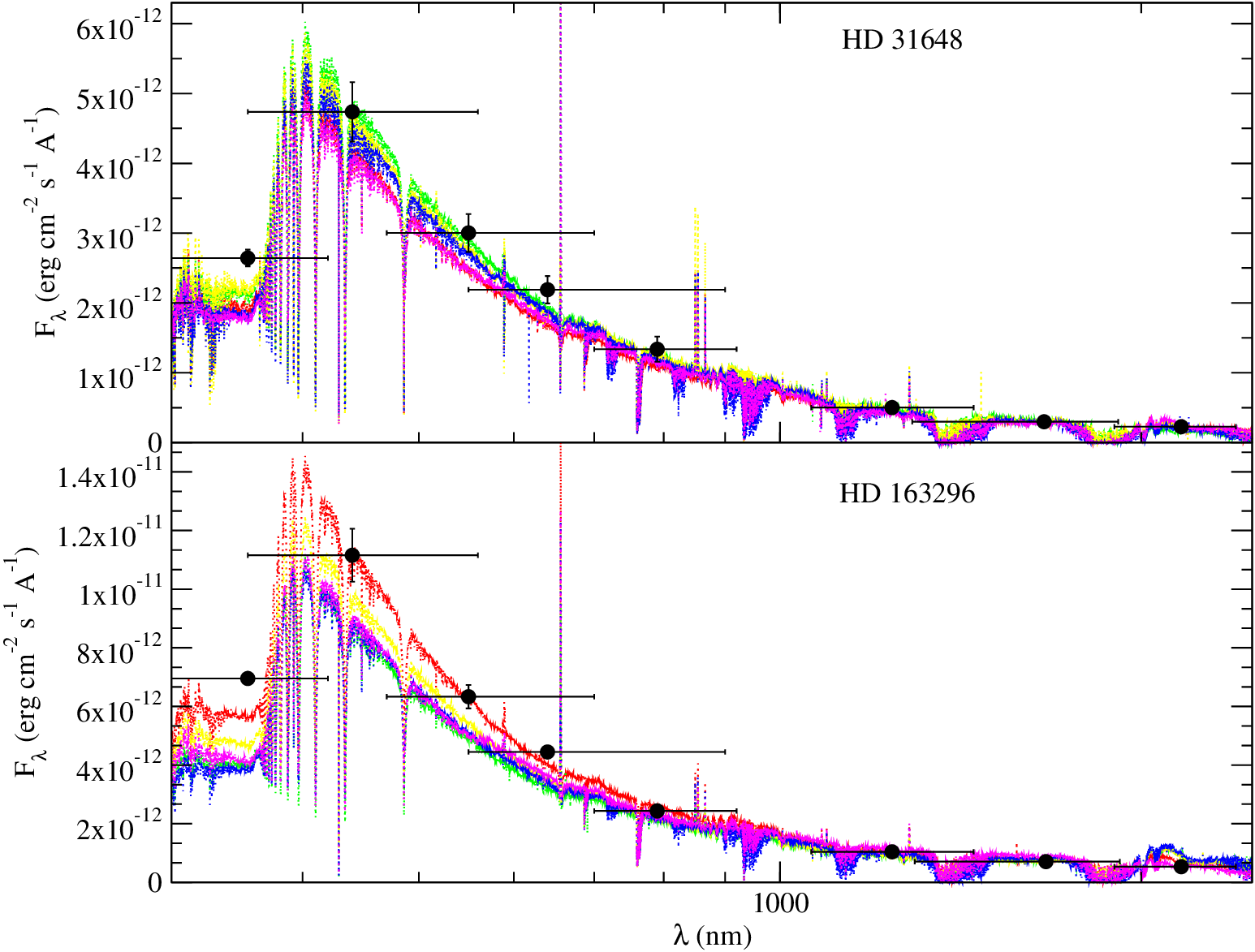}
\caption{X-Shooter spectra of HD 31648 and HD 163296 in red, green, yellow, blue and magenta, representing observing runs A, B, C, D and E, respectively. Selected broadband photometry and uncertainties taken from the literature are plotted with solid circles and vertical bars. The horizontal bars represent each total filter passband.}
\label{Figure:fluxcalib}
\end{figure*}

HD 31648 and HD 163296 are HAe stars \citep[spectral types A5 and A1, ages of $\sim$ 7 Myr and 5 Myr, respectively;][]{Mora01,Montesinos09} with no reported stellar companions in the literature. We assume the parameters derived by \citet{Montesinos09} for the surface temperature, stellar mass and radius (surface gravity), as well as for the distances (T$_{*}$= 8250 $K$, 9250 $K$; M$_{*}$ = 2.0 M$_{\odot}$, 2.2 M$_{\odot}$; R$_*$ = 2.3 R$_{\odot}$, 2.3 R$_{\odot}$; d = 146 pc, 130 pc; where the first and second values refer to HD31648 and HD 163296, respectively). The stellar luminosities derived in that work are 22 L$_{\odot}$ (HD 31648) and 34.5L$_{\odot}$ (HD 163296). Both objects show continuum and line variations \citep{Sitko08}, and have magnetically driven accretion signatures \citep{Swartz05, Hubrig06}. We obtained multi-epoch X-Shooter spectra of both objects. Table \ref{Table:log} shows the log of the observations. A total of 10 science spectra were taken, 5 spectra per star on a timescale from days to months. X-Shooter covers a wide wavelength region throughout its three arms: UVB (311--558 nm), VIS (558--1012 nm) and NIR (1012-2475 nm). The slit widths were 1.6, 0.9 and 0.9 $\arcsec$, which translates into spectral resolutions of 3300, 8800 and 5600 for the UVB, VIS and NIR arms. The spectra were bias and flat field-corrected, flux and wavelength calibrated from standard procedures using the X-Shooter pipeline v1.5.0. \citep{Modigliani10}. Atmospheric emission was subtracted from an ABBA dithering pattern. Additional telluric absorption correction was applied for specific lines in the NIR arm (Pa$\gamma$ and Br$\gamma$; see Sect. \ref{Subsection: varlines}), by using telluric standards that were observed before and after the target stars. Flux calibration was tested from low-resolution spectra -i.e. obtained with the widest slit available for all arms (5$\arcsec$)-, taken consecutively for the target stars and several telluric standard main-sequence objects with similar air masses. Synthetic photometry obtained by convolving the spectra with broadband  $UBVRIJHK$ photometric filters were compared to published values for the telluric standard stars. In this way, the typical uncertainty is in-between 0.05 and 0.2 magnitudes (flux relative errors $\sim$ 5 and 25 $\%$), where the specific uncertainties for each star depend on the X-Shooter arm and observing run (see below). Further accuracy could be obtained from spectro-photometric standards, which were not provided. Synthetic photometry extracted from our spectra is included in Table \ref{Table:log}, which is consistent with previous measurements \citep{dewinter01,Oudmaijer01,Eiroa01}. Flux calibrated spectra are shown in Fig. \ref{Figure:fluxcalib}. Values from broadband photometry published in the literature \citep{Oudmaijer01,Eiroa01} are over plotted for comparison.\\

\begin{table*}
\centering
\caption{Log of the observations and synthetic photometry}
\label{Table:log}
\begin{tabular}{lrrrrr}
\hline\hline
Star&Run A&Run B&Run C&Run D&Run E\\
\hline
HD 31648 &23/10/11&27/10/11&5/11/11&23/02/12&28/02/12\\
\hline
U&8.16 $\pm$ 0.09&8.1 $\pm$ 0.2&8.0 $\pm$ 0.1&8.2 $\pm$ 0.1&8.2 $\pm$ 0.2\\
B&8.18 $\pm$ 0.09&8.0 $\pm$ 0.2&8.1 $\pm$ 0.1&8.1 $\pm$ 0.1&8.2 $\pm$ 0.2\\
V&7.91 $\pm$ 0.09&7.8 $\pm$ 0.2&7.8 $\pm$ 0.1&7.8 $\pm$ 0.1&7.9 $\pm$ 0.2\\
R&7.83 $\pm$ 0.05&7.7 $\pm$ 0.2&7.72 $\pm$ 0.09&7.73 $\pm$ 0.09&7.8 $\pm$ 0.2\\
I&7.65 $\pm$ 0.05&7.5 $\pm$ 0.2&7.49 $\pm$ 0.09&7.57 $\pm$ 0.09&7.6 $\pm$ 0.2\\
J&7.19 $\pm$ 0.09&7.0 $\pm$ 0.2&7.1 $\pm$ 0.1&7.2 $\pm$ 0.2&7.3 $\pm$ 0.1\\
H&6.46 $\pm$ 0.09&6.4 $\pm$ 0.2&6.4 $\pm$ 0.1&6.5 $\pm$ 0.2&6.6 $\pm$ 0.1\\
K&5.70 $\pm$ 0.09&6.1 $\pm$ 0.2&5.7 $\pm$ 0.1&5.8 $\pm$ 0.2&5.6 $\pm$ 0.1\\
\hline
HD 163296&12/10/11&14/10/11&16/10/11&24/03/12&17/05/12\\
\hline
U&7.01 $\pm$ 0.09&7.4 $\pm$ 0.2&7.2 $\pm$ 0.1&7.4 $\pm$ 0.1&7.3 $\pm$ 0.2\\
B&7.08 $\pm$ 0.09&7.4 $\pm$ 0.2&7.2 $\pm$ 0.1&7.4 $\pm$ 0.1&7.3 $\pm$ 0.2\\
V&6.90 $\pm$ 0.09&7.2 $\pm$ 0.2&7.1 $\pm$ 0.1&7.1 $\pm$ 0.1&7.1 $\pm$ 0.2\\
R&6.88 $\pm$ 0.05&7.1 $\pm$ 0.2&7.03 $\pm$ 0.09&7.06 $\pm$ 0.09&7.0 $\pm$ 0.2\\
I&6.74 $\pm$ 0.05&7.0 $\pm$ 0.2&6.91 $\pm$ 0.09&6.92 $\pm$ 0.09&6.9 $\pm$ 0.2\\
J&6.21 $\pm$ 0.05&6.4 $\pm$ 0.2&6.37 $\pm$ 0.09&6.32 $\pm$ 0.09&6.1 $\pm$ 0.2\\
H&5.39 $\pm$ 0.05&5.5 $\pm$ 0.2&5.50 $\pm$ 0.09&5.37 $\pm$ 0.09&5.2 $\pm$ 0.2\\
K&4.52 $\pm$ 0.05&4.3 $\pm$ 0.2&4.30 $\pm$ 0.09&4.18 $\pm$ 0.09&4.7 $\pm$ 0.2\\
\hline
\end{tabular}
\begin{minipage}{179mm}

  \textbf{Notes.} Observing dates (dd/mm/yy) and synthetic photometry, in magnitudes, obtained by convolving Johnson-Cousins $UBVRI$ \citep{Bessell79} and $JHK$ \citep{Cohen03} filter passbands with each flux-calibrated spectrum.
\end{minipage} 
\end{table*} 

\section{Observational Results}
\label{Section:results}

\subsection{Balmer excesses}
\label{Subsection: varbalmer}

The Balmer excess is defined from the $UB$ Johnson's photometric bands as $\Delta$D$_B$ = $(U-B)$$^{phot}$ -- $(U-B)$$^{dered}$, where $(U-B)$$^{phot}$ is the photospheric color, and $(U-B)$$^{dered}$ the observed, dereddened color \citep{Mendi11b}. This can be derived from the extinction (in magnitudes), which for a given wavelength can be expressed from that in the $V$ band, or from the $B-V$ color excess, A$_{\lambda}$ = A$_V$ (k$_{\lambda}$/k$_V$) = R$_{\rm V}$$E(B-V)$(k$_{\lambda}$/k$_V$). The opacity ratio k$_{\lambda}$/k$_V$ is known from an extinction law, and R$_{\rm V}$ is the total-to-selective extinction ratio. The X-Shooter UVB spectra were used to measure continuum excesses in the Balmer region, along the different observing runs. Therefore, the expression for the Balmer excess has been converted into fluxes. If the observed ones are normalized to the photospheric emission at the $B$ band (hereafter represented by the superscript ``norm, B''), then:

\begin{eqnarray}
\label{Eq:Balmer}
\Delta D_B = 2.5\log \left(\frac{\rm F_{U}^{norm, B}}{\rm F_{U}^{phot}}\right)+\nonumber \\ 
+ 2.5R_{\rm V}\left(\frac{k_{U}}{k_{V}}-\frac{k_{B}}{k_{V}}\right)\log \left(\frac{\rm F_{V}^{norm, B}}{\rm F_{V}^{phot}}\right),
\end{eqnarray}


which is the expression that is used in this work. The first term in Eq. \ref{Eq:Balmer} is essentially the same one as used by \citet{DonBrit11}, with F$_{U}^{norm, B}$/F$_{U}^{phot}$ the ratio between the normalized and the photospheric fluxes, measured at wavelengths corresponding to the $U$-band. The second term reflects the contribution of extinction, and can be estimated from an extinction law and from the ratio between the normalized and photospheric fluxes at wavelengths corresponding to the $V$ band. Apart from being independent of absolute flux calibration \citep{DonBrit11}, the main advantage of the above expression for $\Delta$D$_B$ is that it accounts for continuum excesses that can be ascribed mainly to hot shocks caused by accretion (Sect. \ref{Subsection: accretion_from_Balmer}), given that the possible contribution of extinction is ruled out by the second term. This contribution is typically an order of magnitude lower than the first term for the stars studied here -their $E(B-V)$ values are low, see Sect. \ref{Subsection: varlines}-, but it should not be neglected for objects with more pronounced extinctions \citep[i.e. UXOr-like, see e.g. ][and references therein]{Grinin94}. It is assumed that the third possible cause argued to be in the origin of photometric variability in PMS stars, photospheric variations caused by cold spots \citep{Herbst94}, is negligible for the stellar temperatures and masses of both stars analyzed. The photospheric fluxes are represented by Kurucz models \citep{Kurucz93} with the corresponding values of T$_{*}$ and log g (Sect. \ref{Sect:observs}). These were derived from the SYNTHE code, re-binning to a resolution similar to the X-Shooter spectra, as they were also used to estimate the photospheric contribution to the spectral lines (Sect. \ref{Subsection: varlines}). Observed fluxes were normalized to the Kurucz ones, making their averaged continuum fluxes in the 400--460 nm band (representing the rough range covered by the $B$ photometric filter) to coincide. The ratios F$_{U}^{norm, B}$/F$_{U}^{phot}$ and F$_{V}^{norm, B}$/F$_{V}^{phot}$ were then derived considering averaged continuum fluxes in the wavelength regions 350--370 nm and 540--560 nm, respectively. Balmer excesses were finally obtained from Eq. \ref{Eq:Balmer}, assuming R$_{\rm V}$ = 5 \citep{Hernandez04}, k$_{U}$/k$_V$ = 1.6, and k$_{B}$/k$_V$ = 1.3 \citep{Rieke85,Robitaille07}. The use of a larger R$_{\rm V}$ value, compared with the typical for the interstellar medium (R$_{\rm V}$ = 3.1), has been extensively justified for HAeBe stars \citep[see e.g.][and references therein]{Herbst82,Gorti93,Hernandez04,Manoj06}. In addition,  interstellar extinction is most likely negligible compared with the circumstellar one for objects located closer than 200 pc \citep{Fitzgerald68}, as it is the case for HD 31648 and HD 163296. A different R$_{\rm V}$ value than assumed here affects the second term in Eq. \ref{Eq:Balmer}. In particular, R$_{\rm V}$ = 3.1 provides Balmer excesses than can be lower by 0.08 magnitudes, being this upper limit obtained from specific observing runs of HD 163296. For simplicity, possible variations of R$_{\rm V}$ for a given star, or strong dependences of the extinction law on R$_{\rm V}$, are not considered for the low-extinctions and the wavelength range covered in this work \citep{Mathis90}.\\

\begin{figure*}
\centering
\begin{tabular}{cc}
\includegraphics[width=87mm,clip=true]{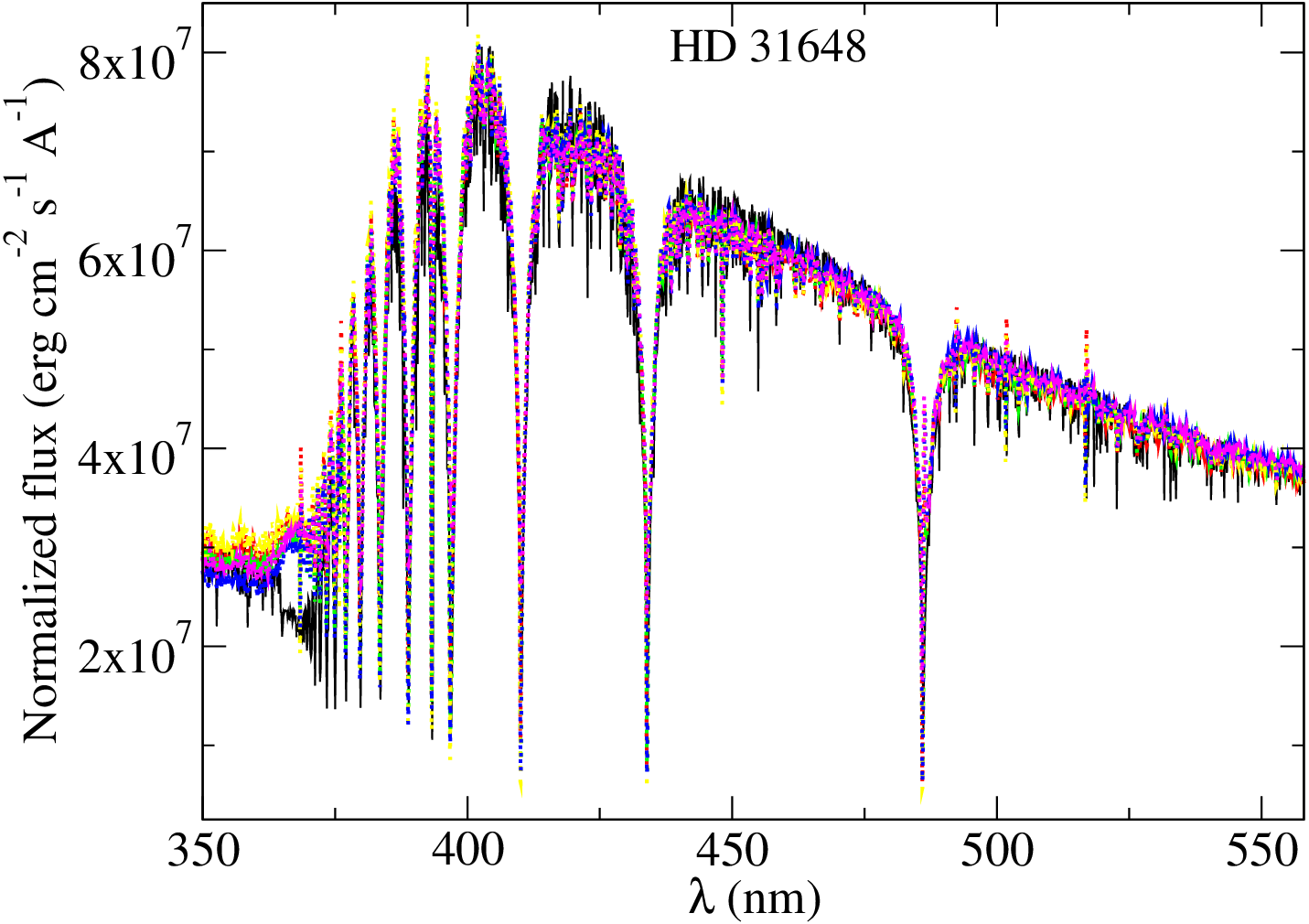} &
\includegraphics[width=87mm,clip=true]{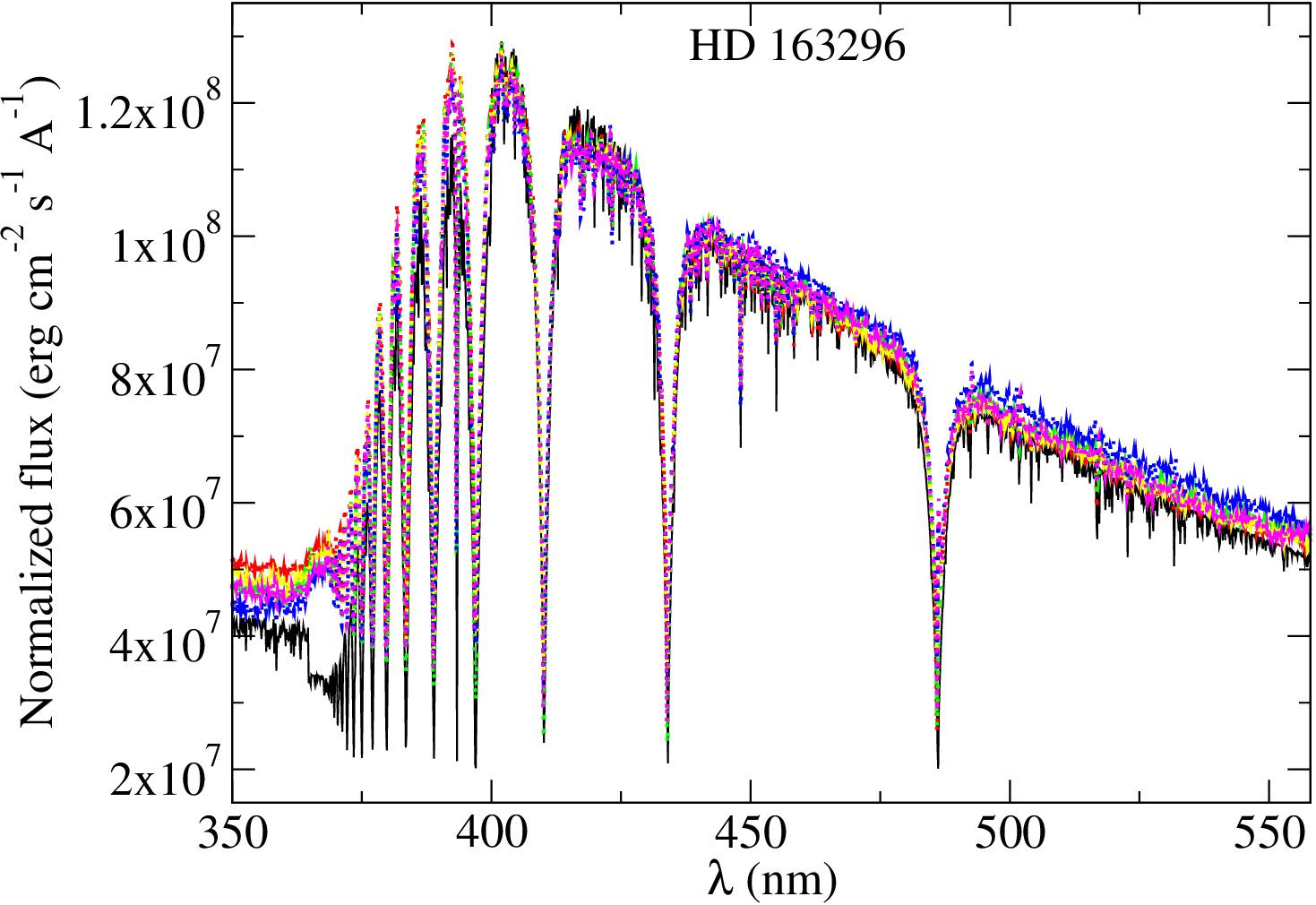} \\
\end{tabular}
\caption{X-Shooter UVB spectra of HD 31648 and HD 163296, normalized to Kurucz photospheric spectra at the stellar surface in the 400--460 nm band. Normalized spectra are plotted in different colors representing different observing runs, as in Fig. \ref{Figure:fluxcalib}. Kurucz spectra are plotted with black solid lines.}
\label{Figure:Balexcess}
\end{figure*}

Figure \ref{Figure:Balexcess} shows the observed spectra for the two stars studied here, normalized to the Kurucz ones at wavelengths corresponding to the $B$ photometric filter. Balmer excesses with respect the photospheric spectra are apparent in the 350--370 nm region, and those can vary from one epoch to another. Table \ref{Table:Balexcess} shows the measured $\Delta$D$_B$ values. Apart from the extinction law, the uncertainty of the Balmer excesses is only dependent of relative flux ratios in wavelengths corresponding to the $U$ and $V$ photometric filters (see Eq. \ref{Eq:Balmer}). In other words, the accuracy of the Balmer excess would only decrease if errors in flux calibration are strongly wavelength-dependent in the 350-560 nm region, which is not the case for our spectra. A conservative uncertainty of 0.01 magnitudes was estimated by measuring -the theoretically null- $\Delta$D$_B$ in consecutive X-Shooter spectra of main sequence stars in different observing dates and with different air masses, for which neither accretion nor changes in extinction were present. The last column in Table \ref{Table:Balexcess} shows the mean Balmer excess and its relative variability, $\sigma$($\Delta$D$_B$)/$<$$\Delta$D$_B$$>$, defined as the ratio between the standard deviation and the average Balmer excess from the different observing runs. While HD 163296 shows the largest mean Balmer excess (0.41 magnitudes), HD 31648 shows the largest Balmer excess variations (up to a factor 2.5 in a few days). It is noted again the importance of the second term in Eq. \ref{Eq:Balmer}: despite of the similar variability shown in the Balmer region of the spectra in both panels of Fig. \ref{Figure:Balexcess}, the contribution of -variable- extinction to this variability is a bit more pronounced in HD 163296 (its normalized flux shows larger differences with respect the Kurucz model at longer wavelengths).  

\begin{table*}
\centering
\renewcommand\arraystretch{1.0}
\renewcommand\tabcolsep{2.5pt}
\caption{Observed Balmer excesses.}
\label{Table:Balexcess}
\begin{tabular}{lcccccc}
\hline\hline
Star&$\Delta$D$_B$ (Run A)&$\Delta$D$_B$ (Run B)&$\Delta$D$_B$ (Run C)&$\Delta$D$_B$ (Run D)&$\Delta$D$_B$ (Run E)&$<$$\Delta$D$_B$$>$\\
 &(mag)&(mag)&(mag)&(mag)&(mag)&(mag)\\
\hline
HD 31648 &0.15&0.12&0.18&0.07&0.18&0.14 [0.33]\\
HD 163296&0.37&0.38&0.41&0.38&0.36&0.38 [0.04]\\
\hline
\end{tabular}

\begin{minipage}{179mm}

  \textbf{Notes.} Observed Balmer excesses are shown in Cols. 2 to 6. The typical uncertainty is 0.01 magnitudes. The last Col. shows the mean Balmer excess and a number in brackets quantifying the relative variability.  
\end{minipage}
\end{table*}

\subsection{Spectral lines}
\label{Subsection: varlines}
\begin{figure*}
\centering
\begin{tabular}{cc}
\includegraphics[width=125mm,clip=true]{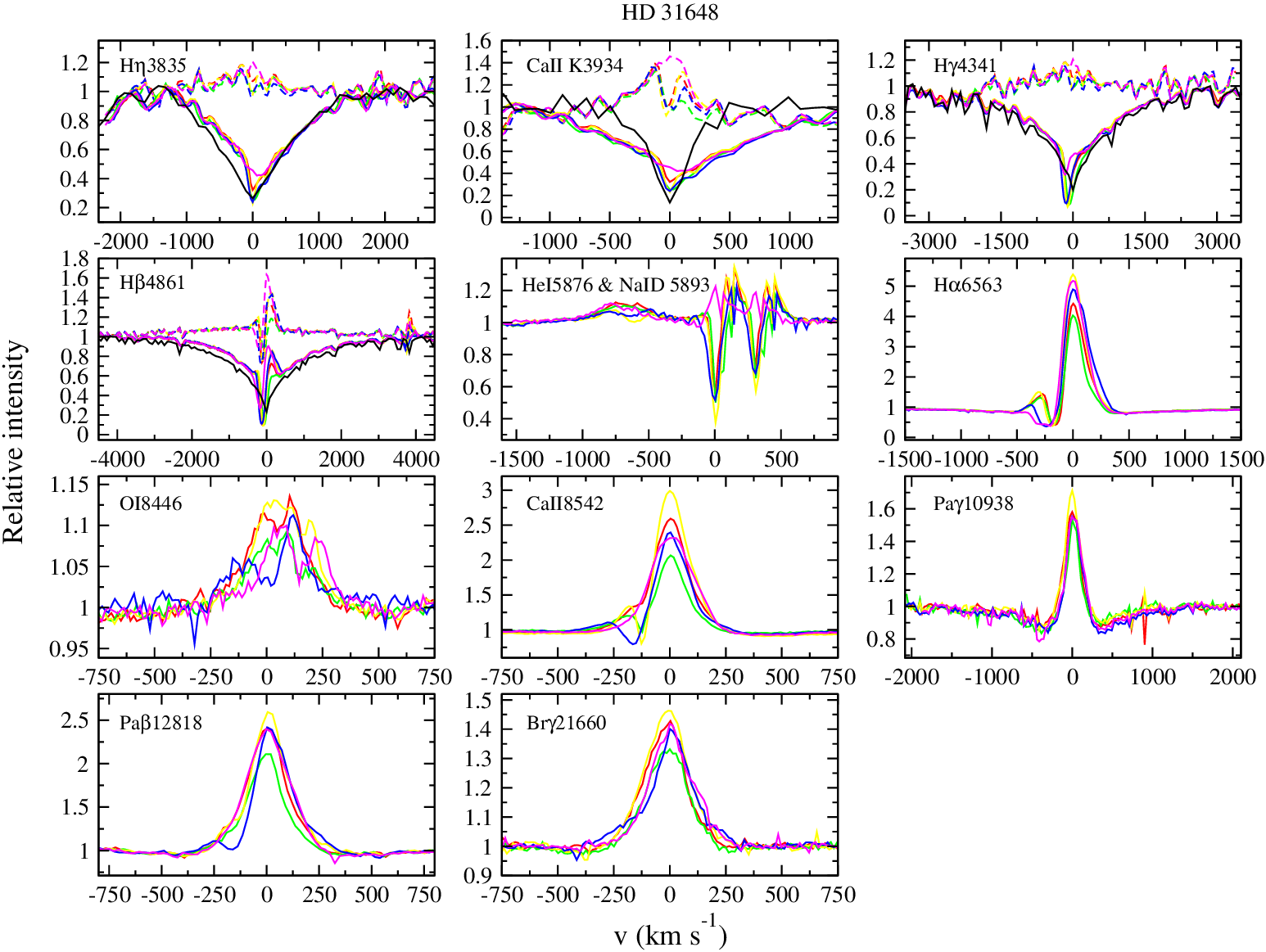}\\
\includegraphics[width=125mm,clip=true]{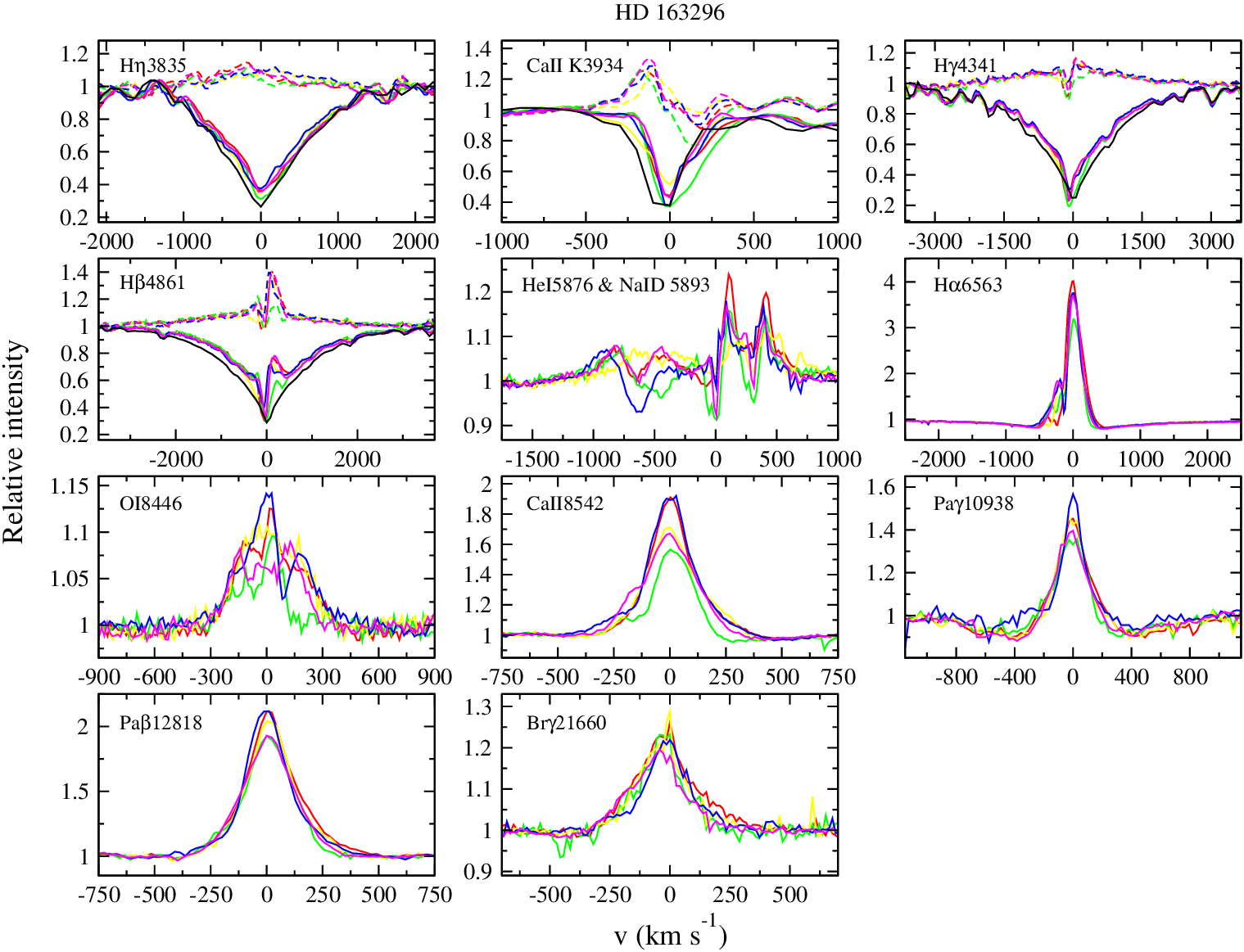} \\
\end{tabular}
\caption{Observed line profiles (normalized to unity) of HD 31648 and HD 163296 are shown with solid lines. Different colors represent different observing runs, as in Fig. \ref{Figure:fluxcalib}. For the UV lines with a purely absorption profile (H$\eta$, \ion{Ca}{2}K, H$\gamma$ and H$\beta$), the emission profiles (dashed lines) that result from the subtraction of the Kurucz synthetic spectra (solid black lines) from the observed ones are also shown (see text).}
\label{Figure:lines}
\end{figure*}

This work focuses on twelve emission lines for which there are previous empirical calibrations with the accretion luminosity in the literature (Sect. \ref{Subsection: accretion_lines}). These lines span over the whole UVB-VIS-NIR wavelength region covered by the X-Shooter spectra. In particular, we focused on transitions of neutral hydrogen, helium, sodium, oxygen, also including transitions of ionized calcium. Figure \ref{Figure:lines} shows the observed line profiles of HD 31648 and HD 163296, along the five observing runs. Forbidden emission lines most probably tracing outflow processes ([\ion{O}{1}], [\ion{S}{2}], [\ion{Fe}{1}]) were not detected in any of the spectra. However, both stars show variable blue-shifted self-absorptions in several lines (H$\alpha$, \ion{Ca}{2}8542, Pa$\beta$), as well as variable red-shifted self-absorptions in others (\ion{Ca}{2}K, H$\beta$, \ion{O}{1}8446). Several works deal with the physical origin of some of these lines \citep[see e.g.][]{Hartmann94,Muzerolle98,Muzerolle01,Muzerolle04,Tambovtseva99,Kurosawa06}, which have been associated mainly with accretion and/or outflow processes. The study of the physical origin of the spectral lines is beyond the scope of this work.\\

The circumstellar contribution to the equivalent width (EW) of a line was determined from EW$^{cs}$ = EW$^{obs}$ - EW$^{phot}$. The observed equivalent width, EW$^{obs}$, was directly measured on the normalized spectra\footnote{The entire line, considering both the emission and absorption components, is taken for this measurement.}, whereas the photospheric absorption, EW$^{phot}$, was estimated from the Kurucz spectra with the corresponding stellar parameters (see Sect. \ref{Subsection: varbalmer}). These profiles were broadened by stellar rotation, asuming projected rotational velocities of 133 and 102 km s$^{-1}$ for HD 163296 and HD 31648, respectively \citep{Mora01}. For the resolution of our X-Shooter spectra, variations of 500 $K$ in the stellar temperature and changes of 0.50 dex in the surface gravity of the Kurucz spectra \citep[uncertainties of $\pm$ 200 $K$ and $\pm$ 0.2 dex were quoted in][]{Montesinos09} translate into typical relative errors of ~ 10 $\%$ and ~ 5 $\%$ in EW$^{phot}$ (the specific values depending on the specific star and line). However, since we are mainly interested on the variability of the lines and the photospheric contribution is expected to be constant, it is assumed that the uncertainty in EW$^{cs}$ is dominated by that in EW$^{obs}$. This was determined for each individual line and observing run by measuring additional EWs from two levels located at $\pm$1.5$\sigma$ from the normalized continuum, fixed at unity. Table \ref{Table:EWs} shows the equivalent widths and the errors estimated in this way. The relative variability, $\sigma$(EW$^{cs}$)/$<$EW$^{cs}$$>$ \citep[see e.g.][]{Mendi11a}, is indicated in those cases where the EW changes above the uncertainties.\\

\begin{table*}
\centering
\renewcommand\arraystretch{1.0}
\renewcommand\tabcolsep{2pt}
\caption{Line equivalent widths}
\label{Table:EWs}
\begin{tabular}{lccccccc}
\hline\hline
& & &HD 31648&&&\\
\hline
     &EW$^{phot}$&EW$^{cs}$&EW$^{cs}$&EW$^{cs}$&EW$^{cs}$&EW$^{cs}$\\
Line &&(run A)&(run B)&(run C)&(run D)&(run E)&$<$EW$^{cs}$$>$\\
     &($\AA$)& ($\AA$)&($\AA$)&($\AA$)&($\AA$)&($\AA$)&($\AA$)\\
\hline
H$\eta$ (3835 \AA{})&10.4&-1.0 $\pm$ 0.1&-1.7 $\pm$ 0.1&-0.7 $\pm$ 0.7&0.1 $\pm$ 0.7&-1.4 $\pm$ 0.8&-1.2 [0.60]\\
\ion{Ca}{2}K (3934 \AA{})&4.5&-1.1 $\pm$ 0.2&-0.2 $\pm$ 0.1&-1.2 $\pm$ 0.1&-0.7 $\pm$ 0.1&-1.9 $\pm$ 0.2&-1.0 [0.65]\\
H$\gamma$ (4341 \AA{})&23.6&-4 $\pm$ 3&-3 $\pm$ 2&-5 $\pm$ 3&-3 $\pm$ 2&-3 $\pm$ 3&-3 $\pm$ 1\\
H$\beta$ (4861 \AA{})&34.5&-18 $\pm$ 2&-17 $\pm$ 2 &-19 $\pm$ 2& -19 $\pm$ 2&-19 $\pm$ 2& -18 $\pm$ 1\\
\ion{He}{1} (5876 \AA{})&0.02&-1.3 $\pm$ 0.4&-1.2 $\pm$ 0.4& -0.8 $\pm$ 0.4& -0.8 $\pm$ 0.4 &-1.0 $\pm$ 0.3& -1.0 $\pm$ 0.2\\
\ion{Na}{1}D$_{2}$ (5890 \AA{})&0.26&-0.57 $\pm$ 0.06&0.09 $\pm$ 0.05&-0.23 $\pm$ 0.08&0.25 $\pm$ 0.06&-0.94 $\pm$ 0.06&-0.58 [0.84]\\
\ion{Na}{1}D$_{1}$ (5896 \AA{})&0.19&-0.60 $\pm$ 0.09&-0.06 $\pm$ 0.06&-0.35 $\pm$ 0.07&-0.07 $\pm$ 0.07&-0.70$\pm$ 0.06& -0.36 [0.80]\\
H$\alpha$ (6563 \AA{})&21.5&-25.6 $\pm$ 0.1& -23.6 $\pm$ 0.1& -28.3 $\pm$ 0.2 & -25.5 $\pm$ 0.2& -25.0 $\pm$ 0.1& -25.6 [0.07]\\
\ion{O}{1} (8446 \AA{})&0.26&-1.3 $\pm$ 0.2&-1.0 $\pm$ 0.2&-1.4 $\pm$ 0.2&-1.1 $\pm$ 0.2&-1.1 $\pm$ 0.2&-1.1 $\pm$ 0.1\\
\ion{Ca}{2} (8542 \AA{})&3.5&-13.9 $\pm$ 0.3&-9.2 $\pm$ 0.2&-15.2 $\pm$ 0.3&-10.7 $\pm$ 0.3&-13.0 $\pm$ 0.3&-12.4 [0.20]\\
Pa$\gamma$ (10938 \AA{})&24.9&-26 $\pm$ 1&-23 $\pm$ 1&-27 $\pm$ 1&-22$\pm$ 1&-24 $\pm$ 1&-25 [0.08]\\
Pa$\beta$ (12818 \AA{})&22.8&-36.0 $\pm$ 0.6&-32.0 $\pm$ 0.5&-37.1 $\pm$ 0.6&-34.6 $\pm$ 0.6&-36.0 $\pm$ 0.6&-35.1 [0.06]\\
Br$\gamma$ (21660 \AA{})&26.3&-33.2 $\pm$ 0.6&-31.3 $\pm$ 0.7&-34.3 $\pm$ 0.6&-32.9 $\pm$ 0.7&-33.1 $\pm$ 0.7&-33.0 [0.03]\\
\hline
& & &HD 163296&&&\\
\hline
     &EW$^{phot}$&EW$^{cs}$&EW$^{cs}$&EW$^{cs}$&EW$^{cs}$&EW$^{cs}$\\
Line &&(run A)&(run B)&(run C)&(run D)&(run E)&$<$EW$^{cs}$$>$\\
     &($\AA$)& ($\AA$)&($\AA$)&($\AA$)&($\AA$)&($\AA$)&($\AA$)\\
\hline
H$\eta$ (3835 \AA{})&10.4&-0.3 $\pm$ 0.4&0.1 $\pm$ 0.4&-0.2 $\pm$ 0.3&-0.4 $\pm$ 0.3&-0.2 $\pm$ 0.4&-0.3 $\pm$ 0.2\\
\ion{Ca}{2}K (3934 \AA{})&2.7&-0.3 $\pm$ 0.1&0.4 $\pm$ 0.1&-0.52 $\pm$ 0.09&-0.50 $\pm$ 0.09&-0.76 $\pm$ 0.08&-0.34 [1.3]\\
H$\gamma$ (4341 \AA{})&23.2&-1 $\pm$ 3&0 $\pm$ 3&-2 $\pm$ 3&-3 $\pm$ 3&-1$ \pm$ 3&-2$\pm$ 1\\
H$\beta$ (4861 \AA{})&29.2&-12 $\pm$ 1&-12.4 $\pm$ 0.8&-12 $\pm$ 1&-13 $\pm$ 1&-12 $\pm$ 2&-12.5 $\pm$ 0.6\\
\ion{He}{1} (5876 \AA{})&0.03&-0.7 $\pm$ 0.1&-0.5 $\pm$ 0.3&-0.8 $\pm$ 0.2&-0.3 $\pm$ 0.3&-0.8 $\pm$ 0.3&-0.7 $\pm$ 0.1\\
\ion{Na}{1}D$_{2}$ (5890 \AA{})&0.16&-0.8 $\pm$ 0.1&-0.34 $\pm$ 0.07&-0.51 $\pm$ 0.07&-0.70 $\pm$ 0.07&-0.53 $\pm$ 0.08&-0.58 [0.33]\\
\ion{Na}{1}D$_{1}$ (5896 \AA{})&0.14&-0.8$\pm$0.1&-0.36$\pm$0.09&-0.62$\pm$0.08&-0.57 $\pm$ 0.07&-0.87 $\pm$ 0.09&-0.64 [0.31]\\
H$\alpha$ (6563 \AA{})&30.2&-35 $\pm$ 4&-30 $\pm$ 4&-33 $\pm$ 3&-34 $\pm$ 3&-33 $\pm$ 4&-33 $\pm$ 2\\
\ion{O}{1} (8446 \AA{})&0.24&-1.3 $\pm$ 0.2&-0.8 $\pm$ 0.1&-1.4 $\pm$ 0.3&-1.4 $\pm$ 0.3&-1.1 $\pm$ 0.2&-1.2 [0.23]\\
\ion{Ca}{2} (8542 \AA{})&3.1&-9.1$\pm$0.4&-5.2$\pm$0.9&-8.5$\pm$0.3&-9.7$\pm$0.4&-8.4$\pm$0.3&-8.2 [0.21]\\
Pa$\gamma$ (10938 \AA{})&23.2&-23 $\pm$ 1&-24 $\pm$ 1&-24 $\pm$ 1& -26.3 $\pm$ 0.6 &-23 $\pm$ 1& -24.0 [0.05]\\
Pa$\beta$ (12818 \AA{})&20.7&-31.3 $\pm$ 0.7&-28.1 $\pm$ 0.7&-28.8 $\pm$ 0.7&-31.1 $\pm$ 0.9&-29.5 $\pm$ 0.7&-29.8 [0.05]\\
Br$\gamma$ (21660 \AA{})&22.2&-27.0 $\pm$ 0.4&-25.6 $\pm$ 0.4&-26.2 $\pm$ 0.6&-25.8 $\pm$ 0.5&-25.7 $\pm$ 0.5& -26.0 [0.02]\\
\hline
\end{tabular}
\begin{minipage}{179mm}

  \textbf{Notes.} Photospheric and circumstellar equivalent widths are shown in Cols. 2 to 7. The last Col. shows the mean equivalent width and its uncertainty from the propagation of the individual ones. For the stars showing both absorptions and emissions, only the values of
the most frequent absorption/emission behavior are considered to derive the mean values. A number in brackets quantifying the relative variability ($\sigma$(EW$^{cs}$)/$<$EW$^{cs}$$>$) is provided when the equivalent width varies above the uncertainties.
\end{minipage}
\end{table*}

Dereddened line luminosities (L$_{\lambda}^{dered}$) were obtained from the values for EW$^{cs}$, the observed continuum fluxes at wavelengths close to the lines (F$_{cont, \lambda}^{obs}$), and from the distances to the stars (d). We used the expression L$_{\lambda}^{dered}$ = 4$\pi$d$^2$ $\times$ F$_{\lambda}^{dered}$, where F$_{\lambda}^{dered}$ = EW$^{cs}$ $\times$ F$_{cont, \lambda}^{dered}$ = EW$^{cs}$ $\times$ F$_{cont, \lambda}^{obs}$ $\times$ 10$^{0.4A_{\lambda}}$. The extinction, A$_{\lambda}$ = R$_{\rm V}$$E(B-V)$(k$_{\lambda}$/k$_V$), was derived assuming again R$_{\rm V}$ = 5, and from the k$_{\lambda}$/k$_V$ values in  \citet{Rieke85} and \citet{Robitaille07}. $(B-V)$ color excesses were obtained subtracting the corresponding magnitudes extracted from the Kurucz spectra from the ones included in Table \ref{Table:log}. These are shown in Table \ref{Table:Ls}, along with the dereddened line luminosities and their uncertainties (which consider those in the EW and in F$_{cont, \lambda}^{obs}$), as well as the relative variability when a line shows luminosity variations above the uncertainties.\\

\begin{table*}
\centering
\renewcommand\arraystretch{0.95}
\renewcommand\tabcolsep{4pt}
\caption{Dereddened line luminosities}
\label{Table:Ls}
\begin{tabular}{lcccccc}
\hline\hline
& & &HD 31648&&&\\
$E(B-V)$     &0.08     &    0.09 &     0.10&    0.13 &     0.09&                      \\
\hline
     &log L/L$_{\odot}$&log L/L$_{\odot}$&log L/L$_{\odot}$&log L/L$_{\odot}$&log L/L$_{\odot}$\\
Line &(run A)&(run B)&(run C)&(run D)&(run E)&log $<$L/L$_{\odot}$$>$\\
\hline
H$\eta$ (3835 \AA{})&-2.3 $\pm$ 0.4&-2.0 $\pm$ 0.2&$<$ -2.07 &$<$-2.38 &-2.1 $\pm$ 0.3&-2.1 $\pm$ 0.2\\
\ion{Ca}{2}K (3934 \AA{})&-2.22 $\pm$ 0.09&-2.9 $\pm$ 0.3&-2.07 $\pm$ 0.06&-2.3 $\pm$ 0.1&-2.0 $\pm$ 0.1&-2.19 [0.57]\\
H$\gamma$ (4341 \AA{})&-1.8 $\pm$ 0.4&-1.8 $\pm$ 0.4&-1.6 $\pm$ 0.3&-1.8 $\pm$ 0.3&$<$ -1.52 &-1.7 $\pm$ 0.2\\
H$\beta$ (4861 \AA{})&-1.21 $\pm$ 0.06&-1.2 $\pm$ 0.1&-1.10 $\pm$ 0.06&-1.03 $\pm$ 0.06&-1.2 $\pm$ 0.1&-1.13 [0.16]\\
\ion{He}{1} (5876 \AA{})&-2.6 $\pm$ 0.1&-2.6 $\pm$ 0.2&-2.7 $\pm$ 0.2&-2.7 $\pm$ 0.2&-2.7 $\pm$ 0.2&-2.6 $\pm$ 0.1\\
\ion{Na}{1}D$_{2}$ (5890 \AA{})&-2.95 $\pm$ 0.05&...&-3.3 $\pm$ 0.2&...&-2.71 $\pm$ 0.09&-2.92 [0.60]\\
\ion{Na}{1}D$_{1}$ (5896 \AA{})&-2.94 $\pm$ 0.07&...&-3.09 $\pm$ 0.09&...&-2.83 $\pm$ 0.09&-2.94 [0.28]\\
H$\alpha$ (6563 \AA{})&-1.43 $\pm$ 0.02&-1.4 $\pm$ 0.1&-1.31 $\pm$ 0.04&-1.31 $\pm$ 0.04&-1.41 $\pm$ 0.09&-1.37 [0.13]\\
\ion{O}{1} (8446 \AA{})&-3.01 $\pm$ 0.07&-3.1 $\pm$ 0.1&-2.89 $\pm$ 0.07&-2.99 $\pm$ 0.09&-3.1 $\pm$ 0.1&-3.00 [0.17]\\
\ion{Ca}{2} (8542 \AA{})&-1.97 $\pm$ 0.02&-2.1 $\pm$ 0.1&-1.86 $\pm$ 0.04&-1.99 $\pm$ 0.04&-1.97 $\pm$ 0.09&-1.97 [0.19]\\
Pa$\gamma$ (10938 \AA{})&-1.96 $\pm$ 0.04&-2.0 $\pm$ 0.1& -1.86 $\pm$ 0.05& -2.0 $\pm$ 0.1&-1.99 $\pm$ 0.05& -1.94 [0.11]\\
Pa$\beta$ (12818 \AA{})&-1.93 $\pm$ 0.04&-1.9 $\pm$ 0.1&-1.87 $\pm$ 0.05&-1.9 $\pm$ 0.1&-1.95 $\pm$ 0.04&-1.91 $\pm$ 0.03\\
Br$\gamma$ (21660 \AA{})&-2.28 $\pm$ 0.04&-2.5 $\pm$ 0.1&-2.29 $\pm$ 0.05&-2.4 $\pm$ 0.1&-2.22 $\pm$ 0.04&-2.32 [0.23]\\
\hline
& & &HD 163296&&&\\
$E(B-V)$     &0.10     &    0.13 &     0.10&    0.16 &     0.15&                      \\
\hline
     &log L/L$_{\odot}$&log L/L$_{\odot}$&log L/L$_{\odot}$&log L/L$_{\odot}$&log L/L$_{\odot}$\\
Line &(run A)&(run B)&(run C)&(run D)&(run E)&log $<$L/L$_{\odot}$$>$\\
\hline
H$\eta$ (3835 \AA{})&$<$ -2.06&$<$ -2.48&$<$ -2.28 &-2.3 $\pm$ 0.4&$<$ -2.27&-2.3 $\pm$ 0.4\\
\ion{Ca}{2}K (3934 \AA{})&-2.4 $\pm$ 0.2&...&-2.18 $\pm$ 0.09&-2.09 $\pm$ 0.09&-1.9 $\pm$ 0.1&-2.11 [0.41]\\
H$\gamma$ (4341 \AA{})&$<$ -1.31&$<$ -1.47&$<$ -1.28&$<$ -1.10& $<$ -1.28&...\\
H$\beta$ (4861 \AA{})&-1.00 $\pm$ 0.05&-1.0 $\pm$ 0.1&-1.06 $\pm$ 0.06&-0.91 $\pm$ 0.05&-1.0 $\pm$ 0.1&-0.99 [0.13]\\
\ion{He}{1} (5876 \AA{})&-2.52 $\pm$ 0.06&-2.8 $\pm$ 0.3&-2.5 $\pm$ 0.1&$<$ -2.57&-2.5 $\pm$ 0.2&-2.6 $\pm$ 0.1\\
\ion{Na}{1}D$_{2}$ (5890 \AA{})&-2.47 $\pm$ 0.06&-2.9 $\pm$ 0.1&-2.75 $\pm$ 0.07&-2.52 $\pm$ 0.06&-2.6 $\pm$ 0.1&-2.63 [0.38]\\
\ion{Na}{1}D$_{1}$ (5896 \AA{})&-2.49 $\pm$ 0.06&-2.9 $\pm$ 0.2&-2.66 $\pm$ 0.07&-2.61 $\pm$ 0.07&-2.4 $\pm$ 0.1&-2.58 [0.37]\\
H$\alpha$ (6563 \AA{})&-0.99 $\pm$ 0.05&-1.1 $\pm$ 0.1&-1.07 $\pm$ 0.06&-0.98 $\pm$ 0.05&-1.0 $\pm$ 0.1&-1.02 $\pm$ 0.04\\
\ion{O}{1} (8446 \AA{})&-2.71 $\pm$ 0.07&-3.0 $\pm$ 0.1&-2.8 $\pm$ 0.1&-2.7 $\pm$ 0.1&-2.8 $\pm$ 0.1&-2.76 [0.26]\\
\ion{Ca}{2} (8542 \AA{})&-1.86 $\pm$ 0.03&-2.2 $\pm$ 0.1&-1.96 $\pm$ 0.04&-1.83 $\pm$ 0.04&-1.88 $\pm$ 0.09&-1.92 [0.26]\\
Pa$\gamma$ (10938 \AA{})&-1.73 $\pm$ 0.03&-1.8 $\pm$ 0.1&-1.79 $\pm$ 0.04&-1.67 $\pm$ 0.04 & -1.66 $\pm$ 0.09&-1.73 [0.15]\\
Pa$\beta$ (12818 \AA{})&-1.71 $\pm$ 0.02&-1.8 $\pm$ 0.1&-1.80 $\pm$ 0.04&-1.69 $\pm$ 0.04&-1.66 $\pm$ 0.09&-1.73 [0.15]\\
Br$\gamma$ (21660 \AA{})&-2.00 $\pm$ 0.02& -1.9 $\pm$ 0.1 & -1.90 $\pm$ 0.04 & -1.84 $\pm$ 0.04 & -2.13 $\pm$ 0.09& -1.94 [0.24]\\
\hline
\end{tabular}
\begin{minipage}{179mm}

  \textbf{Notes.} Emission line luminosities in Cols. 2 to 6 are dereddened using the color excesses indicated on the top of each column (in magnitudes), and a distance of 146 and 130 pc for HD 31648 and HD 163296, respectively. The last Col. shows the mean line luminosity and its uncertainty from the propagation of the individual ones. A number in brackets quantifying the relative variability ($\sigma$(L)/$<$L$>$) is provided when the line luminosity varies above the uncertainties. Upper limits are not considered to derive mean and relative variability values. Blank spaces refer to lines shown in absorption.
\end{minipage}
\end{table*}

Although line luminosities tend to be larger for HD 163296, the variability of the lines, both in terms of the EWs and the line luminosities, are comparable for HD 31648 and HD 163296.\\

\section{Analysis and discussion}
\label{Sect:analysis}

\subsection{Accretion rates from Balmer excesses}
\label{Subsection: accretion_from_Balmer}
It is assumed that accretion is magnetically channeled for the two stars studied in this paper. Therefore, the observed Balmer excesses and their variations (Sect. \ref{Subsection: varbalmer}) are analyzed in terms of a MA shock model. We follow the nomenclature and methodology described in \citet{Mendi11b}, where several assumptions of the MA shock model in \citet{CalvetGullbrig98} and \citet{Muzerolle04} were applied to HAeBe stars. According to that, the total flux per wavelength unit emerging from the star ($F_{\lambda}$) is composed of the emission from the naked photosphere ($F_{\lambda}^{phot}$, represented by a Kurucz synthetic spectrum) plus the accretion contribution: 
\begin{equation}
\label{Eq:totflux}
F_{\lambda} = fF_{\lambda}^{col} + (1 - f)F_{\lambda}^{phot}
\end{equation}
$f$ being the filling factor that reflects the stellar surface coverage of the accretion columns, and $F$$_{\lambda}^{col}$ the flux from the column. This parameter is modeled as a blackbody at a temperature:
\begin{equation}
\label{Eq:F}
T_{col} = [\curf/\sigma + \rm T_{*}^4]^{1/4} 
\end{equation}
with $\sigma$ the Stefan-Boltzmann constant and $\curf$ the inward flux of energy carried by the accretion columns, which ranges typically between 10$^{11}$ and 10$^{12}$ erg cm$^{-2}$ s$^{-1}$ \citep{Muzerolle04}. The filling factor can be estimated from: 
\begin{equation}
\label{Eq:condition2}
f = \left(1-\frac{\rm R_*}{\it R_i}\right)\frac{\rm {GM_{*}}\it \dot{M}_{\rm acc}}{4 \pi \curf \rm R_*^3},
\end{equation}

\begin{table}
\centering
\renewcommand\tabcolsep{1.5pt}
\caption{Fixed model parameters}
\label{Table:input}
\begin{tabular}{lccc}
\hline\hline
Parameter&Symbol (units)&HD 31648&HD 163296\\
\hline
stellar temperature &T$_*$ ($K$)&8250&9250\\
stellar mass&M$_*$ (M$_{\odot}$)&2.0&2.2\\
stellar radius&R$_*$ (R$_{\odot}$)&2.3&2.3\\
accretion column flux&$\curf$ (erg cm$^{-2}$ s$^{-1}$)&10$^{12}$&10$^{12}$\\
disk truncation radius&$R$$_{i}$ (R$_*$)&2.5&2.2\\
\hline
\end{tabular}
\end{table}

with $R$$_{i}$ the disk truncation radius.\\

\begin{figure}
\centering
 \includegraphics[width=85mm,clip=true]{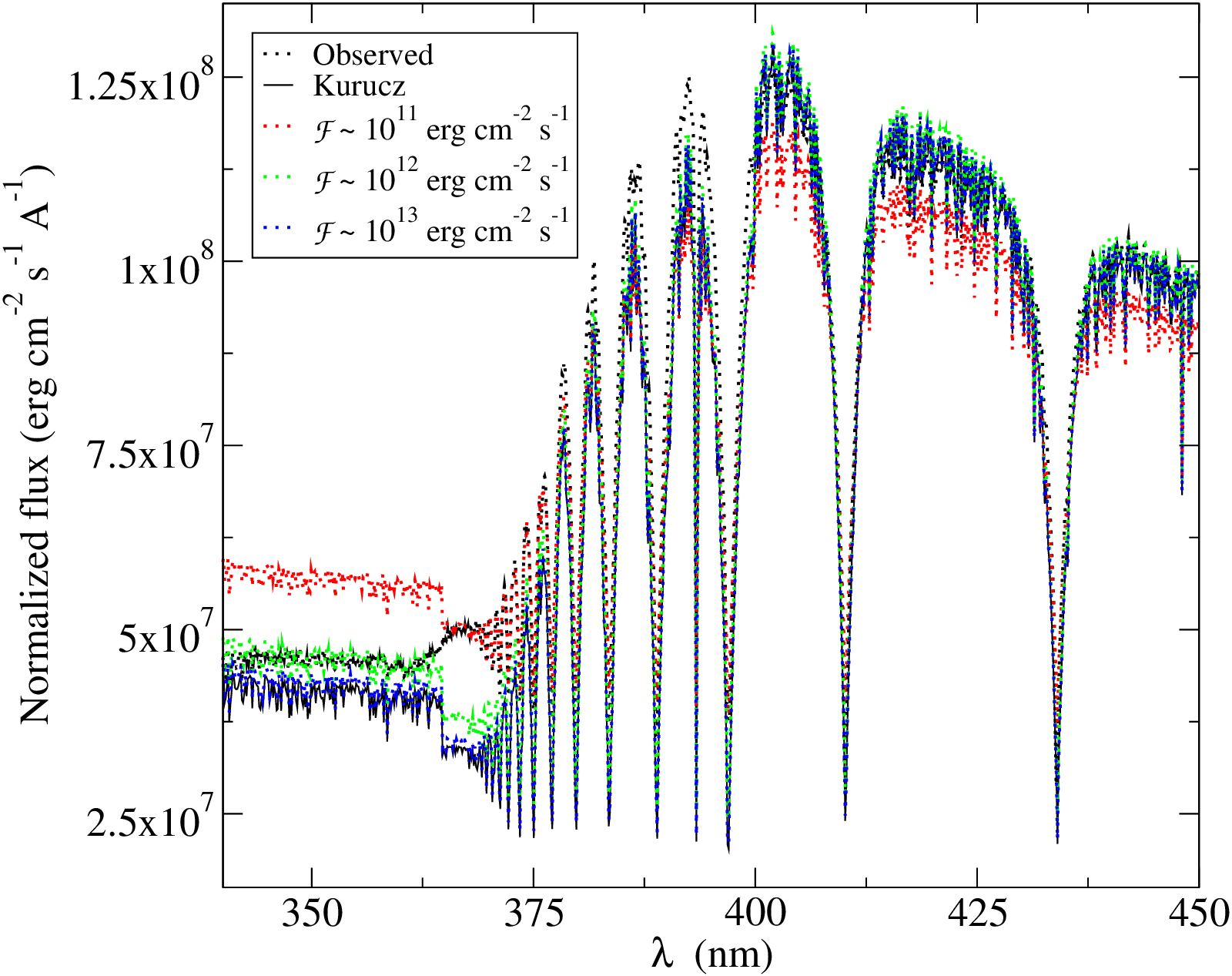}
\caption{Average UV spectrum of HD 163296 (black dotted line) normalized to the kurucz synthetic spectrum (black solid line) at the $B$ band. The coloured dotted lines refer to modelled spectra with fixed mass accretion rate (10$^{-7}$ M$_{\odot}$ yr$^{-1}$), stellar parameters and disk truncation radius (see text), but different values for the inward flux energy, $\curf$, as indicated in the legend. Those values are associated to different blackbody temperatures and filling factors by Eqs. \ref{Eq:F} and \ref{Eq:condition2}: $\sim$ 9700 $K$ (f $\sim$ 20 $\%$), 12500 $K$ (f $\sim$ 2 $\%$) and 25000 $K$ (f $\sim$ 0.1 $\%$) for the red, green and blue lines, respectively.}
\label{Figure:accretion_hd163296}
\end{figure}

\begin{figure}
\centering
 \includegraphics[width=85mm,clip=true]{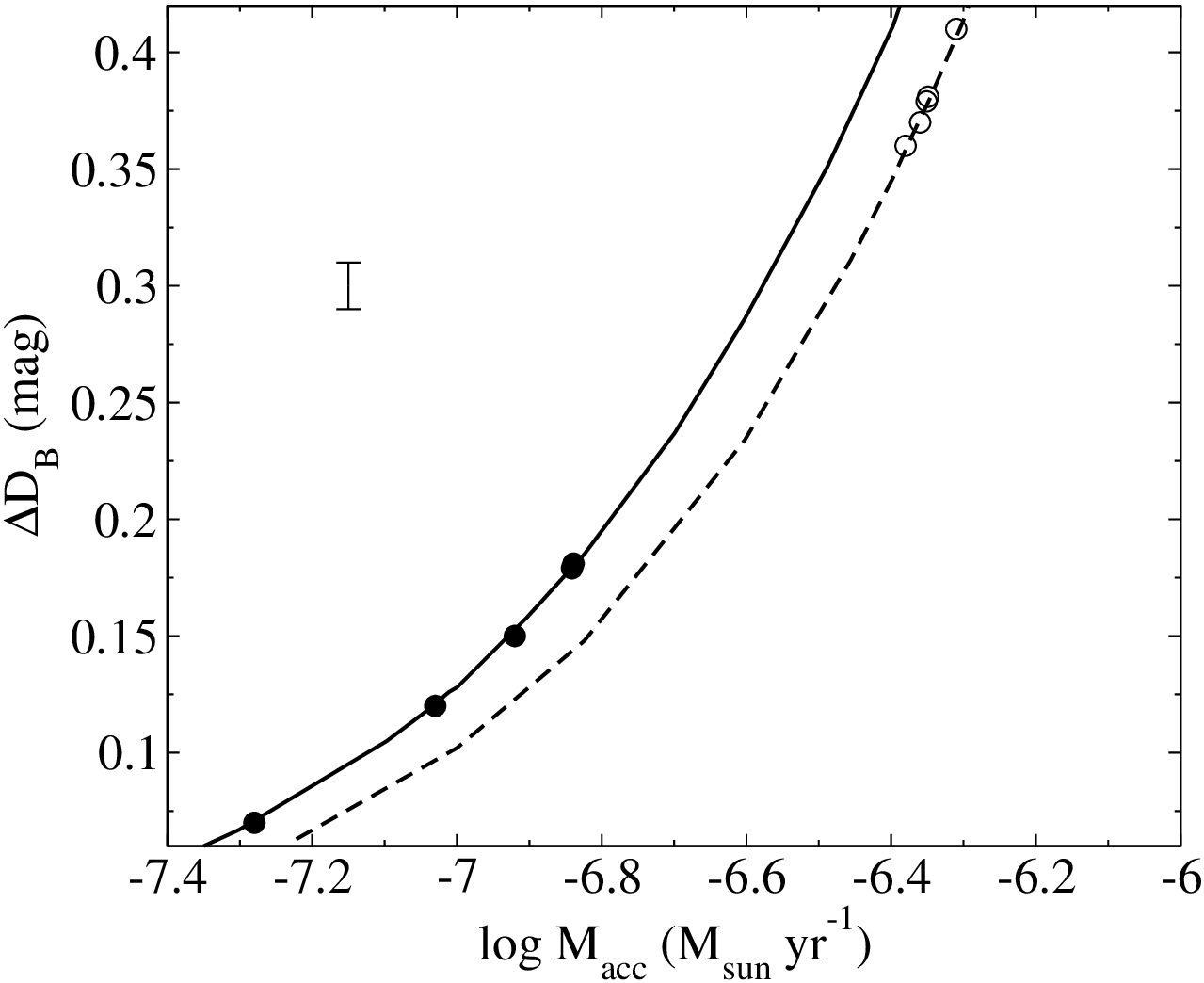}
\caption{Calibration between the Balmer excess and the mass accretion rate for HD 31648 (solid line) and HD 163296 (dashed line). The observed Balmer excesses are over plotted with filled and open circles. The uncertainty in the observed Balmer excesses is represented by the black vertical bar.}
\label{Figure:calibration_Bal_ac}
\end{figure}

Table \ref{Table:input} summarizes the fixed input parameters that are used in this work. Figure \ref{Figure:accretion_hd163296} shows the average UV spectrum of HD 163296 normalized to the Kurucz photospheric spectrum in the $B$ band. For a typical accretion rate of $\sim$ 10$^{-7}$ M$_{\odot}$ yr$^{-1}$, fixed stellar parameters and disk truncation radius (see below), different values of $\curf$ provide different blackbody temperatures (Eq. \ref{Eq:F}) and filling factors (Eq. \ref{Eq:condition2}). The corresponding total fluxes (Eq. \ref{Eq:totflux}) reproduce the observed spectra with different degrees of accuracy. For $\curf$ $\sim$ 10$^{11}$ erg cm$^{-2}$ s$^{-1}$, the only part of the observed spectrum that can be reproduced is the bump between 365 and 370 nm, mismatching at shorter and longer wavelenghts ($\lambda$ $>$ 380 nm). Values of $\curf$ $\sim$ 10$^{12}$ erg cm$^{-2}$ s$^{-1}$ are able to reasonably reproduce both the short and long wavelength regions. Larger values ($\curf$ $\sim$ 10$^{13}$ erg cm$^{-2}$ s$^{-1}$) are not able to reproduce any part of the observed Balmer excesses. As it is described in the following, our modelling aims to reproduce mainly the strength of the observed Balmer excess, instead of its overall shape. We assume a ``classical'' aproach using a single blackbody obtained by fixing $\curf$ to 10$^{12}$ erg cm$^{-2}$ s$^{-1}$, which not only provides reasonable agreement with the observed excesses and filling factors \citep{Valenti93}, but is in turn the typical value for HAe stars \citep{Muzerolle04}. More complex models using several values of $\curf$ have been used to simultaneously reproduce wider wavelength ranges covering the near-UV and optical regions of T Tauri stars \citep[see e.g.][]{Ingleby13}. Disk truncation radii assumed in this work (Table \ref{Table:input}) are also typical \citep{Muzerolle04}, being lower than the corresponding co-rotation radii \citep{Shu94} by taking into account the projected rotational velocities of the stars \citep{Mendi11b}. We refer the reader to this paper for further details on the shock modeling and its dependences on the stellar parameters. Although already discussed there, further analysis on the influence of the adopted values for $\curf$ and R$_i$ is included below. Total fluxes were synthesized from Eq. \ref{Eq:totflux} for each star, considering different values for $\dot{M}_{\rm acc}$. Modeled Balmer excesses were then derived by normalizing F$_{\lambda}$ to F$_{\lambda}^{phot}$ in the the 400--460 nm band, and applying the same procedure as described for the observed Balmer excesses (Sect. \ref{Subsection: varbalmer}). In this way, couples of values ($\dot{M}_{\rm acc}$, $\Delta$D$_B$) were obtained for each star, which were used to construct the curves represented in Fig. \ref{Figure:calibration_Bal_ac}. These $\Delta$D$_B$--$\dot{M}_{\rm acc}$ calibrations were used to assign a mass accretion rate to each observed Balmer excess. Accretion luminosities were then derived considering $L$$_{\rm acc}$ = GM$_{*}$$\dot{M}_{\rm acc}$/R$_*$. The model parameters that best reproduce the values of the observed Balmer excesses are shown in Table \ref{Table:output}. The uncertainties for $\dot{M}_{\rm acc}$ were estimated considering the typical uncertainty for the observed Balmer excess ($\pm$ 0.01 magnitudes, see Sect. \ref{Subsection: varbalmer}) in the $\Delta$D$_B$--$\dot{M}_{\rm acc}$ calibration (Fig. \ref{Figure:calibration_Bal_ac}). Errors in M$_{*}$/R$_*$ ratios \citep{Montesinos09} add another 0.01 dex uncertainty to the accretion luminosities \citep[see][and the caption of Table \ref{Table:output}]{Mendi11b}.\\

\begin{table}
\centering
\caption{Accretion parameters from the observed Balmer excesses}
\label{Table:output}
\begin{tabular}{lcccc}
\hline\hline
Star&Run&log $\dot{M}_{\rm acc}$&log $L$$_{\rm acc}$&f\\
& & [M$_{\odot}$ yr$^{-1}$]&[L$_{\odot}$]& (\%)\\
\hline
 &A&-6.92&0.51&2.3\\
 &B&-7.03&0.40&1.8\\
HD 31648&C&-6.84&0.60&2.8\\
(T$_{col}$ = 12215 $K$) &D&-7.28&0.15&1.0\\
 &E&-6.84&0.60&2.8\\
\hline
 &A&-6.36&1.11&8.5\\
 &B&-6.35&1.13&8.8\\
HD 163296&C&-6.31&1.17&9.7\\
(T$_{col}$ = 12570 $K$) &D&-6.35&1.13&8.8\\
 &E&-6.38&1.10&8.2\\
\hline
\end{tabular}
\begin{minipage}{85mm}

  \textbf{Notes.} Uncertainties for log $\dot{M}_{\rm acc}$ and log $L$$_{\rm acc}$ are $\pm$ 0.03 dex and $\pm$ 0.04 dex (for HD 31648); $\pm$ 0.02 dex and $\pm$ 0.03 dex (for HD 163296).  
\end{minipage}
\end{table}

From the method described before, the mean mass accretion rate for HD 31648 was 1.11 $\times$ 10$^{-7}$ M$_{\odot}$ yr$^{-1}$. The accretion rate changes are above the uncertainties, between 5.24 $\times$ 10$^{-8}$ M$_{\odot}$ yr$^{-1}$ and 1.46 $\times$ 10$^{-7}$ M$_{\odot}$ yr$^{-1}$. This represents a mass accretion rate variation of almost 0.5 dex. The accretion luminosity showed an average value of 3.03 L$_{\odot}$, and changes by a factor of almost 3; from 1.43 to 3.97 L$_{\odot}$. This represents accretion luminosity variations between 6.5 and 18 $\%$ of the stellar one. Regarding HD 163296, the mass accretion rate was typically larger -$\sim$ 4.50 $\times$ 10$^{-7}$ M$_{\odot}$ yr$^{-1}$- and roughly constant -variations lower than 0.1 dex-. The accretion luminosity was typically $\sim$ 40 $\%$ of the stellar one. Despite of this high ratio, significant veiling in photospheric absorption lines is not detected, which is consistent with a temperature of the accretion shocks similar to the effective temperature of the star \citep{Muzerolle04}.\\ 

The above discussion assumes that the fraction of the star covered by the shocks, $f$, is the only model parameter that varies along the observations. This assumption provides upper limits for the accretion rate changes. In other words, more complex scenarios assuming additional variations in $\curf$ and $R$$_{i}$, would explain a given Balmer excess change from a lower accretion rate variation than provided here. As an example, in the case of HD 31648 a change of 20$\%$ in $\curf$ and $R$$_{i}$ would account for Balmer excess variations of 0.2 and 0.4 magnitudes, respectively, for a fixed accretion rate of 1.11 $\times$ 10$^{-7}$ M$_{\odot}$ yr$^{-1}$. Further analysis in this sense is out of the scope of this work. However, it should be mentioned that changes in the disk truncation radius are considered in 3D simulations of MA \citep{Kulkarni13}. In addition, given that there are observed variations in the ratio of the Balmer lines (which is a probe of the density $\rho$), especially in the case of HD31648, it may be likely that the density of material in the columns is varying.  Given that  $\curf$ $\propto$ $\rho$$v$$^{3}$, with $v$ the velocity of the infalling material \citep[see e.g.][]{CalvetGullbrig98}, a change in the Balmer ratio could be indicating variations in $\curf$.\\
\\
\\
\subsection{Accretion rates and line luminosities}
\label{Subsection: accretion_lines}

The accretion luminosities derived from the Balmer excesses are plotted against the dereddened line luminosities in the different panels of Fig \ref{Figure:accretion_line}. Empirical calibrations relating both parameters were taken from the literature (see below), and are over plotted with solid lines. These have the form log L$_{acc}$/L$_{\odot}$ = a log L$_{line}$/L$_{\odot}$ + b, a and b being constants that depend on the spectral line. These relations were calibrated by relating primary accretion signatures (i.e. line veiling or continuum excess) with the line luminosities, and show a typical maximum scatter of  $\sim$ $\pm$ 1 dex, which is represented by the dashed lines. The calibration obtained from HAeBe stars in \citet{Mendi11b} is over plotted in the H$\alpha$ panel. This is very similar to the corresponding relation for lower mass T Tauri stars and brown dwarfs \citep{Dahm08,Herczeg08}. The relation derived by \citet{Calvet04} \citep[see also][]{DonBrit11,Mendi11b} is used for the Br$\gamma$ panel. That was based on intermediate-mass TTs (F, G spectral types and masses up to 4 M$_{\odot}$). For the remaining lines there are no L$_{acc}$--L$_{line}$ relations calibrated from HAeBes. The empirical relations for the \ion{He}{1}5876, \ion{Ca}{2}8542, and  Pa$\beta$ are from \citet{Dahm08}, based on objects with masses up to 2M$_{\odot}$. The calibrations with the rest of the lines were derived from objects with masses below 1M$_{\odot}$, and were taken from \citet{Herczeg08} (H$\eta$, \ion{Ca}{2}K, H$\gamma$, \ion{Na}{1}D, \ion{O}{1}8446), \citet{Gatti08} (Pa$\gamma$) and \citet{Fang09} (H$\beta$).\\

\begin{figure*}
\centering
 \includegraphics[width=179mm,clip=true]{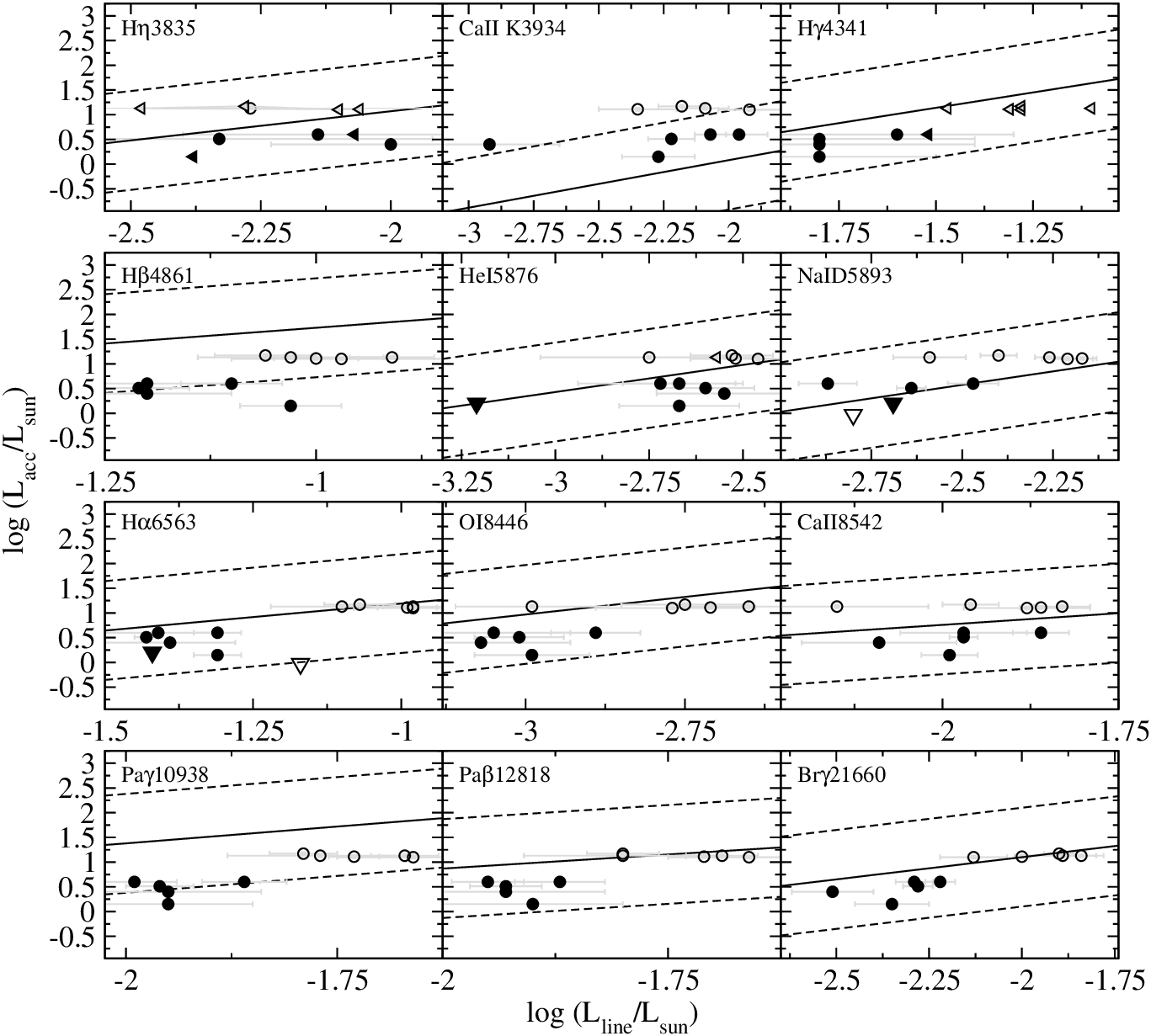}
\caption{Accretion luminosities from the Balmer excess versus emission line luminosities for HD 31648 (filled symbols) and HD 163296 (open symbols). Left-triangles represent upper limits for the line luminosities. Upper limits for the accretion luminosity obtained $\sim$ 15 years ago (see text) are plotted with big triangles (filled and open for HD 31648 and HD 163296, respectively), with respect contemporaneous \ion{He}{1}5876, \ion{Na}{1}D and H$\alpha$ luminosities. Empirical calibrations from previous works \citep{Calvet04,Dahm08,Gatti08,Herczeg08,Fang09,Mendi11b} and $\pm$ 1 dex uncertainties are indicated with solid and dashed lines, respectively.}
\label{Figure:accretion_line}
\end{figure*} 

The accretion and line luminosities plotted in Fig \ref{Figure:accretion_line} fall within the range expected from previous L$_{acc}$--L$_{line}$ calibrations. This suggests that these relations, mostly derived from low-mass stars, can be extrapolated to provide accretion rate estimates for HAe stars. However, the fact that HD 31648 and HD 163296 show different Balmer excess -accretion rate- variability (Sects. \ref{Subsection: varbalmer} and \ref{Subsection: accretion_from_Balmer}) but comparable line luminosities changes (Sect. \ref{Subsection: varlines}) constitutes a first hint suggesting that both variations are not directly related. In order to provide a more quantitative description, we computed the accretion luminosities from the above referred empirical calibrations with the spectral lines. Given that the scatter of these calibrations can reach $\pm$ 1 dex for a given value of a line luminosity, it is unrealistic to expect that they could trace accretion rate variations lower than 0.5 dex, as the ones reported in this work. However, it is possible to compare the accretion variability with the variations from the spectroscopic tracers if residuals with respect the averages are used \citep[e.g.][]{Pogodin12}. Fig. \ref{Figure:Laccevol} shows the difference between single-epoch accretion luminosities derived from the calibrations with each emission line and the average accretion luminosity obtained from the same line. Error bars consider those in the line luminosities. The differences between the single-epoch and the mean accretion luminosities derived from the Balmer excesses are also included in Fig. \ref{Figure:Laccevol} (squares), from which we extract the following:

\begin{figure*}
\centering
 \includegraphics[width=179mm,clip=true]{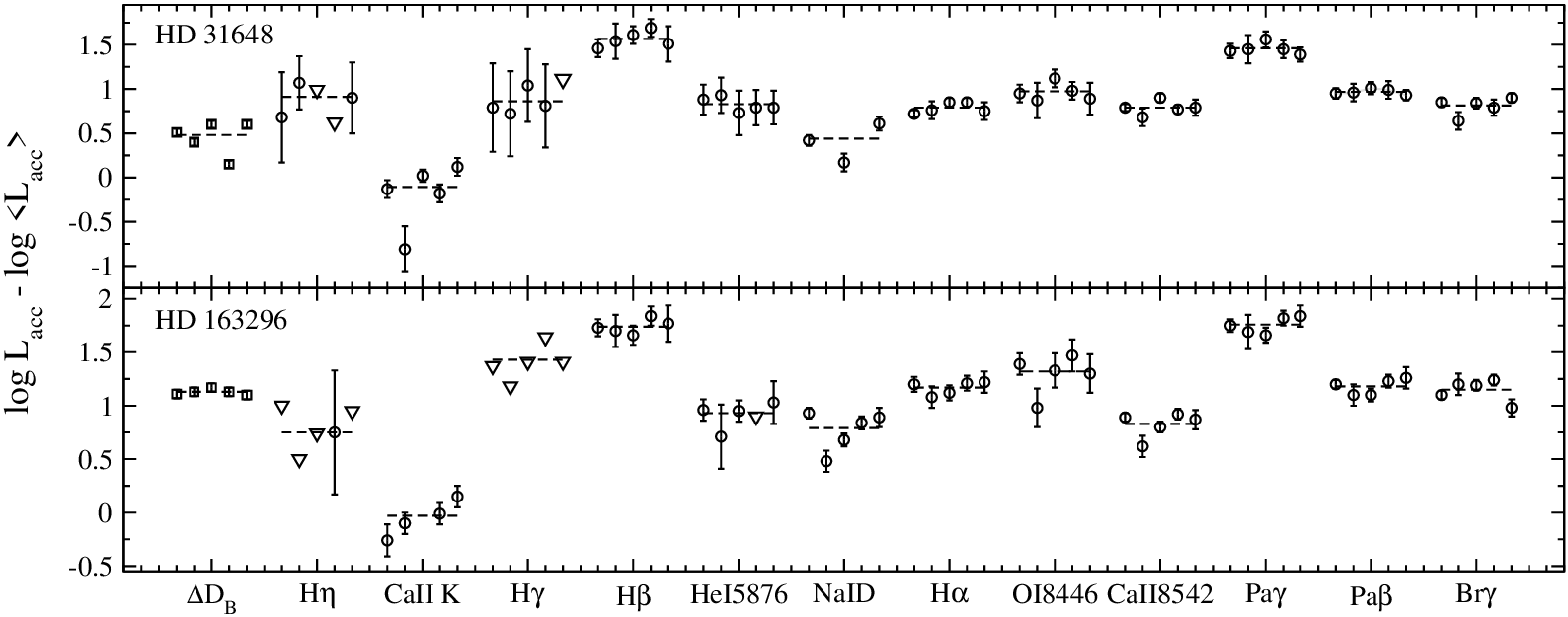}
\caption{Residuals for each single epoch accretion luminosity derived from the indicated tracer with respect the mean accretion luminosity from the same tracer (horizontal dashed lines). Squares refer to the values derived from the Balmer excesses, and triangles represent upper limits. These were not considered to derive the mean values, except for the H$\gamma$ line of HD 163296. For each accretion tracer, the observing runs are ordered chronologically from left to right, spaced by an arbitrary scale.}
\label{Figure:Laccevol}
\end{figure*}


\begin{itemize}
 \item The mean accretion luminosity from the Balmer excess agrees with that derived from previous empirical calibrations within $\pm$ 0.5 dex, for most spectroscopic lines. The only exceptions are the \ion{Ca}{2}K, Pa$\gamma$ and H$\beta$ lines, that provide mean accretion luminosities that can differ up to $\sim$ $\pm$ 1 dex. Not surprisingly, the latter calibrations were derived from objects with stellar masses well below the ones covered in this work \citep{Gatti08,Herczeg08,Fang09}. 
 \item The largest residual for the Balmer excess differs from that for the spectral lines, which in turn provide different values depending on the line considered. For instance, the \ion{Na}{1}D line suggests accretion luminosity variations up to $\sim$ 0.30 dex larger than the largest change observed from the Balmer excess in HD 163296. However, the same line provides a similar upper limit for the accretion luminosity changes of HD 31648. On the contrary, the highest variability from the Pa$\beta$ line roughly coincides with that from the Balmer excess for HD 163296, whereas for HD 31648 suggests accretion rate variations up to $\sim$ 0.30 dex lower. The largest residual from the \ion{Ca}{2}K line is $\sim$ 0.40 and $\sim$ 0.20 dex higher than the corresponding from the Balmer excess, for HD 31648 and HD 163296, respectively.      
 \item In spite of the last point, the residuals derived from some lines show a similar evolution pattern than that from the Balmer excess, which could be supporting a direct link between the physical origin of the lines and the accretion process. In particular, the residuals from the \ion{Ca}{2}K, \ion{Ca}{2}8542 and Br$\gamma$ lines increase/decrease when those from the Balmer excess increase/decrease, for HD 31648. The Br$\gamma$ line also shows a similar behavior for HD 163296. In addition, although different from the residuals from the Balmer excess, the variability behavior is roughly similar for several other lines. For instance, the similar behavior of the H$\beta$, H$\alpha$, Pa$\beta$ and Pa$\gamma$ lines could be suggesting a common physical origin explaining their variability, not necessarily connected with the accretion flow, or connected but with a certain time-delay \citep[see][and references therein]{Dupree13}. However, it is noted that these lines provide roughly constant estimates, considering the error bars, and more accurate measurements become necessary.
\end{itemize}

In summary, our results provide evidence that the L$_{acc}$--L$_{line}$ calibrations can be extrapolated to HAe stars to estimate typical, mean accretion rates, although the relations for some specific lines should most probably include minor modifications in order to achieve a $\pm$ 0.5 dex accuracy. However, the L$_{acc}$--L$_{line}$ calibrations cannot generally be used to quantify the accretion rate variability. In addition, the analysis of accretion variability in terms of residuals provides evolution patterns that, if confirmed by follow-up observations, could be useful to understand the influence of accretion on the origin of the lines, and possible physical relations between the lines themselves. In this respect, accurate spectro-photometry should be performed in order to detect the smallest possible changes in the line luminosities. \\

\subsection{Variability on a longer timescale}
\label{Subsection: var_lont}
    
The methodology to obtain the calibration between the Balmer excess and the accretion rate described in Sect. \ref{Subsection: accretion_from_Balmer} is the same one than that applied in \citet{Mendi11b}, as well as the stellar and modeling parameters used for HD 31648 and HD 163296. Therefore, a direct comparison is feasible. $UB$ photometry taken more than 15 years ago by the EXPORT consortium \citep{Eiroa00}, was analyzed in that paper to provide accretion luminosities of HAeBe stars. Only upper limits could be derived for the two objects studied in this work, due to the lower accuracy of broad band photometry compared with intermediate-resolution spectra. \citet{Mendi11b} provided L$_{acc}$ $<$ 1.58 L$_{\odot}$ (HD 31648) and L$_{acc}$ $<$ 0.93 L$_{\odot}$ (HD 163296). The accretion luminosity of HD 31648 during the X-Shooter observing run D is consistent with that work. Contemporaneous H$\alpha$ and \ion{Na}{1}D emission line luminosities were also similar to the ones obtained here, although the \ion{He}{1}5876 luminosity was lower in the EXPORT campaigns by $\sim$ 0.5 dex. In the case of HD 163296, the averaged accretion luminosity inferred here is at least 1 dex larger than in the EXPORT campaigns. This strong change in the accretion luminosity from the Balmer excess is accompanied by changes in all optical line luminosities analyzed from the EXPORT spectra. The \ion{He}{1}5876 line was then dominated by absorption whereas it is shown in emission in the X-Shooter data. The \ion{Na}{1}D emission lines increased their luminosity by $\sim$ 0.5 dex from the EXPORT to the X-Shooter campaigns. Similarly, the H$\alpha$ EW changed from an averaged value of $\sim$ -23 $\AA$ to -30 $\AA$ in the current X-Shooter data, implying a line luminosity increase of almost 0.2 dex. Unfortunately, the EXPORT spectra analyzed in \citet{Mendi11b} did not covered the X-Shooter UVB and NIR regions, so it is not possible to explore the variations for the lines in those wavelengths. EXPORT results are over plotted in Fig. \ref{Figure:accretion_line} for comparison. The analysis in a longer timescale suggests similar conclusions than those obtained from the study of the X-Shooter spectra in the previous section, at least for the optical lines: the L$_{acc}$--L$_{line}$ calibrations are useful to estimate typical accretion rates with an accuracy of 0.5--1 dex, but are not reliable to accurately quantify accretion rate variability.\\     

\section{Summary and conclusions}
\label{Sect:conclusions}
We presented five-epoch X-Shooter spectra of the prototypical Herbig Ae stars HD 31648 and HD 163296. The strength of the excess shown in the Balmer region of the spectra, and its variations, have been reproduced from magnetospheric accretion shock modeling. The accretion rate of HD 31648 varied in-between 5.24 $\times$ 10$^{-8}$ and 1.46 $\times$ 10$^{-7}$ M$_{\odot}$ s$^{-1}$ on timescales of days to months, whereas that for HD 163296 remained roughly constant, with a typical value of 4.50 $\times$ 10$^{-7}$ M$_{\odot}$ yr$^{-1}$. Higher accretion rate variations, exceeding 1 dex, are found for HD 163296 on timescales of several years. The mean accretion rates derived from the Balmer excess are consistent with those inferred from the empirical calibrations with twelve UV, optical and NIR emission line luminosities, previously derived from lower-mass stars. This demonstrates that those calibrations can be used also for HAe stars. However, the variability of the accretion rate from the Balmer excess is not generally reflected by that from the line luminosities, suggesting that the empirical calibrations are not useful to derive accurate accretion rate variations.\\

Our results suggest that follow-up multi-epoch observations of these and other pre-main sequence stars will be useful to further constrain the underlying origin of the empirical calibrations with the line luminosities, and the origin of the lines themselves.\\

\acknowledgments

The authors thank the anonymous referee for his/her useful comments on the original manuscript, which helped us to improve the paper.\\
Based on observations made with ESO Telescopes at the La Silla Paranal Observatory
under programme ID 088.C-0218.\\
S.D. Brittain acknowledges support for this work from the National Science Foundation under
grant number AST-0954811\\
C. Eiroa, G. Meeus and B. Montesinos are supported by AYA 2011-26202.\\
G. Meeus is supported by RYC-2011-07920.\\



{\it Facilities:} \facility{VLT:Kueyen}



\end{document}